%% file: main.tex
\documentclass{article}
\usepackage{spconf}
\ninept 
\newcommand{\spaceup}{\vspace{-2.3mm}}

\input{header.tex}
\usepackage[utf8]{inputenc}
\usepackage{fullpage}
\usepackage{mathcomSTEv4}

\newtheorem*{remark}{Remark}

\title{Distributed Convex Optimization With Limited Communications}

\name{Milind Rao$^{\dagger}$ \qquad Stefano Rini$^{\star}$ \qquad Andrea Goldsmith$^{\dagger}$}

\address{$^{\dagger}$ Electrical Engineering, Stanford University, Stanford, CA \\
$^{\star}$ National Chiao Tung University, Taiwan \\
\texttt{ \{milind,andreag\}@stanford.edu, stefano@nctu.edu.tw}}
\date{}

\begin{document}
\maketitle

\begin{abstract}
In this paper, a distributed convex optimization algorithm, termed  \emph{distributed coordinate dual averaging} (DCDA) algorithm, is proposed. 
The DCDA algorithm addresses the scenario of a large distributed optimization problem with limited communication among nodes in the network. 
Currently known distributed subgradient methods, such as the distributed dual averaging  or the distributed  alternating direction method of multipliers algorithms, assume that nodes can exchange messages of large cardinality.
Such network communication capabilities are not valid in many scenarios of practical relevance. 
In the DCDA algorithm, on the other hand, communication of each coordinate of the optimization variable is restricted over time.
For the proposed algorithm, we bound the rate of convergence under different communication protocols and network architectures.
%
We also consider the extensions to the case of imperfect gradient knowledge and the case in which transmitted messages are corrupted by additive noise or are quantized.
Relevant numerical simulations are  also provided.


\end{abstract}

\spaceup
\section{Introduction}
\label{sec:introp}
\spaceup
%
With the emergence in recent years of big data paradigms, decentralized optimization algorithms have received considerables interest in the literature.
%
%
%
%
A distributed optimization problem of particular relevance is the one in which the global objective function is obtained as the sum of a local convex functions.
%
%
Originally considered by Tsitsiklis et al. \cite{tsitsiklis1984problems}, this problem 
is broadly referred to as the consensus problem. 
%
%
A number of distributed subgradient methods have been proposed to solve the consensus problem such as 
the distributed dual averaging 
and the distributed  alternating direction method of multipliers algorithms.
%
A distributed subgradient (DSG) algorithm for the consensus problem is initially proposed in \cite{nedic2009distributed},  building upon consensus algorithms for computing the exact averages of initial values at the agents \cite{olfati2004consensus}.
In the algorithm of \cite{nedic2009distributed}, each node updates its estimate using a linear combination of the estimates of its neighbors and the gradient of its local function. 
In the literature, a number of variations of this  algorithm have been considered, such as continuous time extensions \cite{wang2011control}, 
networks with link failures \cite{lobel2011distributed}, and  quantized communication \cite{yi2014quantized}.
Another interesting variation of the DSG algorithm of \cite{nedic2009distributed} is the coordinate descent method in which, in order to reduce the dimension of the messages sent across a network, only one coordinate of the optimal solution is communicated at each time instant.
For this problem, Liu et al \cite{liu2015asynchronous} analyze an asynchronous distributed coordinate descent algorithm. 
Inspired by Nesterov's dual averaging algorithm \cite{nesterov2009primal}, Duchi et al. \cite{duchi2012dual} prose the distributed dual averaging (DDA) algorithm for the consensus problem.
In this algorithm, each node maintains an estimate for a dual variable by averaging the estimates of its neighbors and adding the gradient. A proximal projection of the dual variable  produces optimization variable.
%
The dual variable is updated similarly to the DSG algorithm, while the dual projection allows to incorporate nonlinear constraints on the solution. 
In \cite{duchi2012dual}, the authors also study the performance of the DDA algorithm in the presence of time varying networks,  communication of gossip protocols, and  stochastic gradients.
%
%
The analysis of the DDA algorithm with delays in the communication  network is performed in  \cite{agarwal2011distributed, tsianos2012distributed}. 
The authors of \cite{tsianos2012communication} study the computation/communication trade-off for the DDA algorithm by considering the case in which communication is subject to a total cost constraint.
%

Another popular class of algorithms to solve distributed constrained convex optimization problems are  distributed  alternating direction method of multipliers
(DADMM) algorithms. 
This class of algorithms was originally proposed in \cite{wei2012distributed}, building upon the ADMM algorithm of \cite{boyd2011distributed}.
The analysis of convergence for this algorithm is performed in \cite{shi2014linear}, while the case of asynchronous communications is studied in \cite{zhang2014asynchronous}.
%
%
%
%
%
%

\smallskip
\noindent
{\bf Contributions:}
%
%
%
Our main contribution is a decentralized distributed protocol that places rate constraints  on the communication between nodes, and yet guarantees convergence to the optimal value at all nodes. 
This algorithm is inspired by the DDA algorithm of \cite{duchi2012dual} and is thus termed the  \emph{distributed coordinate dual averaging} (DCDA) algorithm.
In the following, we derive the convergence of DCDA algorithm for different communication protocols and network architectures. 
Additionally, we study the behavior of the algorithm in the scenario of a stochastic gradient, noisy and quantized communication.


%
%
%
%
%
%
%
%
%
%
%

\spaceup
\section{Problem formulation}
\label{sec:Problem formulation}
\spaceup 

We study the distributed optimization problem in which the minimum of a function is to be computed when factors of this functions are distributed across a network  subject to communication constraints. 
Consider the $n$-nodes undirected graph $G=(V,E)$, $V=[1:n]$, and $E \subset V\times V$ in which each the node $V_i$ 
is associated  the function $f_i: \Rbb^d \goes \Rbb$.
Each function $f_i$ is a factor of the linear combination
\ea{
f(x)=\sum_{i=1}^n f_i(x),
\label{eq:global function}
}
for $x \in \Xcal$ with $\mathcal{X}$ closed and convex. We assume that each $f_i(x)$ is convex and L-Lipschitz with respect to a norm $\| \cdot \|$ or $|f_i(x)-f_i(y)| \leq  L \|x-y\|, ~~  x, y \in \Xcal$. The Lipschitz condition implies that for any $x \in \Xcal$  and any subgradient $g_i \in \partial f_i(x)$, we have $\|g_i\|_* \leq L$.
At each time instant $t \in \Nbb$, the node $V_i$ maintains an estimate $x_i(t)$ of the value $x^*$ which attains the minimum of the function $F(x)$ in \eqref{eq:global function}.
%
The node $V_i$ is able to communicate to the node $V_j$ at the time instant $t$  if the two nodes are connected by an edge $E$ in $G$. Let $A$ be the symmetric incidence matrix of $G$.\footnote{That is, $A_{ij}$ is non-zero only if nodes $i$ and $j$ are neighbors.} Some examples of a network are:

\noindent
$-$ {\bf fully-connected network:}  in which $A = \mathbf{1}\mathbf{1}^\intercal - \mathbf{I}$,

\noindent
$-$ {\bf random network:} in which two nodes are connected with probability $p$,

\noindent
$-$ {\bf ring network:} in which $A_{ij, i\neq j}= 1 ~ \mathrm{iff} ~  |i-j| \mod n \leq l$ for some $l \in \Nbb$. In this configuration, nodes are arranged in a circle and a node is connected to $l$ neighbors on either sides. 

\smallskip

\noindent
Upon receiving a message from its neighboring nodes, each node $V_i$ updates its estimate of the minimum value, $x_i(t)$.
In the distributed optimization problem, the goal is to determine a set of communication strategies and estimate update rules such that each $x_i$ converges 
to $x^*$ as time grows to infinity.

In the following, given the time sequence $c(t) \in \Rbb^n$, we will denote the time and space average as  $\ch(t)$ and $\co(t)$ respectively, i.e.
\ean{
\ch_i(t) = \frac{1}{t}\sum_{t'=1}^t x_i(t'), \quad \co(t)=\f 1 n \sum_{i=1}^n c_i(t).
}
%

\smallskip
\noindent
{\bf The DCDA algorithm:}
In  the DCDA algorithm, each node $V_i$ maintains both an estimate of the optimization variable, $x_i(t)$, and its dual variable, $z_i(t)$. 
At each time instant, both the primal variable and the dual variables are updated according to the message received from the neighboring nodes and the sub-differential of the objective function $f_i$ in the primal estimate $x_i$, $g_i(t)$. 
%
%
More precisely, the  communication and update rules are as follows.
At each iteration, each node $i$ broadcasts a subset of its $d$ coordinates of the dual variable $z(t)$ to a subset of of its neighbors. For instance, node $i$ broadcasts coordinate $k$ to neighbors $N^k(i)$. 
The update of the dual variable is a component wise update
\begin{align}
[z_i(t+1)]_k &= \sum_{j \in N^k(i)} P^k_{ij}(t) [z_j(t)]_k + [g_i(t)]_k \quad \forall \ k,
\label{eq:DCDA dual update}
\end{align}
\spaceup

\noindent
where $P^k(t)$ is a doubly stochastic matrix  and where $P^k_{ij}>0$ if and only if
%
$A_{ij}>0$ and the node $j$ is broadcasting the set of coordinates $k$ to node $i$. 
In the following, we consider three different policies for the selection of the coordinate $k$ broadcasted by the nodes:

\noindent
$-$ {\bf static sharing scheme:} at each time instant, nodes transmit the same coordinates to their neighbors, corresponding to $P^k(t)=P^k$ for some fixed $P^k$. 

\noindent
$-$ {\bf round robin scheme:} in which the $k^{\rm th}$ coordinate is shared every $\pi$ time instances, corresponding to $P^k(t)=P_\pi$ when $t= n \pi +k$ for some $n \in \Nbb$, else $P^k(t)=\mb{I}$.

\noindent
$-$ {\bf randomized scheme:} in which nodes randomly and uniformly select the coordinate to be transmitted in each time instant.

Note that the stated sharing scheme with $P^k=P$ corresponds to the DDA of \cite{duchi2012dual}: this  corresponds to the case when nodes broadcast their entire dual variable to their neighbors.
Also note that, given a symmetric adjacency network $A^k(t)$ for coordinate $k$ at time $t$, we can obtain the doubly stochastic matrix $P^k(t)$ as
\ean{
D^k(t) = \mr{diag}(A^k(t)\mathbf{1}),\ \ P^k(t) = \mathbf{I} - \frac{D^k(t)-A^k(t)}{\max_i D^k(t)_{ii}+1}.
}

At each time instant, the primal variable $x_i(t+1)$ is computed from $z_i(t+1)$  as:
\ea{
x_i(t+1) &= \Pi_{\psi,\alpha(t)} (z_i(t+1)) \label{eq:update_x}
}
The function $\Pi_{\psi,\alpha(t)}$ is  type of non-linear proximal projection and is used to stabilize estimates of the primal variable and ensure that optimization constraints are satisfied.  It is defined as
\ea{
\Pi_{\psi,\alpha(t)} (z_i(t)) =\argmin_x \langle x, z_i(t)\rangle + \frac{1}{\alpha(t)} \psi(x).
}
the $\{\al(t)\}_{t=0}^{\infty}$ is a non-increasing sequence of positive step-sizes which typically scales as $1 /\sqrt{t}$.  Also, $\psi: \ \Rbb^d \goes \Rbb$ is a \emph{proximal function},  that is assumed to be 1-strongly convex with respect to norm $\| \cdot \|$.
and positive defined. 
Examples of a proximal function include:

\noindent
$-$ {\bf squared proximal function}: $\psi(x) = \frac{1}{2}\|x\|_2^2$ is 1-strongly convex with respect to the $\ell_2$-norm.

\noindent
$-$ {\bf entropic proximal function}: $\psi(x) = \sum_{k=1}^d x_i\log x_i - x_i$ is 1-strongly convex with respect to the $\ell_1$-norm. 

%
%
\smallskip
\noindent
The performance of the DCDA algorithm is studied in terms of the convergence to zero of the term $f(\hat{x}_i(T))-f(x^*)$.  

Finally, we consider three extensions of the DCDA algorithm: 

\noindent
$-$ {\bf stochastic gradient:} in which the objective function subgradient is not exactly known at each node,

\noindent
$-$ {\bf noisy communication:} in which transmissions are corrupted by additive noise,

\noindent
$-$ {\bf quantized communications:}  in which transmissions are quantized before communication.
%
%

\spaceup
\section{Main results}
\label{sec:Main results}
\spaceup
The main results of the paper consists of the characterization of the DCDA convergence rate for different coordinate selection policies and communication networks.

\begin{thm}
\label{thm:basic_all}
Let the sequences $\{x_i(t)\}_{t=0}^{\infty}$ and $\{z_i(t)\}_{t=0}^{\infty}$ be generated
by the updates \eqref{eq:update_x} and \eqref{eq:DCDA dual update} with step size sequence $\{ \al(t)\}_{t=0}^{\infty}$. 
Then for any $x^* \in \Xcal$ and for each node $i \in V$, the DCDA algorithm is such that  
\ea{
& f(\hat{x}_i(T)) - f(x^*) \leq \frac{\psi(x^*)}{T \alpha(t)}+  \frac{1}{T}\sum_{t=1}^T  \alpha(t-1)\|\bar{g}(t)\|_*^2  
\label{eq:basic_all}\\
%
&  + \f {2L}{n T} \sum_{t=1}^T \sum_{j=1}^n \al(t) ||\zo(t)-z_j||_*+ \f L T \sum_{t=1}^T \al(t)||\zo(t)-z_j||_*.
\nonumber 
}
\end{thm}
The result in Th. \ref{thm:basic_all} is substantially equivalent to the  result, of \cite[Th. 1]{duchi2012dual}. 
The first two terms in \eqref{eq:basic_all} are optimization error terms common to sub-gradient algorithms while the last two are penalties incurred due to having different estimates at different nodes in the network or the penalty from consensus. 
%
%
The result in Th. \ref{thm:basic_all} can be further developed for specific communication protocols.

\begin{lem}{\bf Static  sharing  scheme:}
\label{lem:Static  sharing  scheme}
For the settings in Th. \ref{thm:basic_all}, the DCDA algorithm under the static coordinate sharing scheme is such that 
\ea{
& f(\hat{x}_i(T)) - f(x^*) \leq  \frac{\psi(x^*) }{T \alpha(T)} 
\label{eq:static sharing}\\
 & + \frac{L^2}{T}\sum_{t=1}^T  4 \alpha(t-1) \lb  \frac{2\min(d,n)\log \sqrt{n}dT}{1-\max_k\sigma_2(P^k)} + 3  \rb. \nonumber 
}
where $\sigma_2(M)$ is the second largest eigenvalue of $M$.
\end{lem}

Lem. \ref{lem:Static  sharing  scheme} implies that for the choice of $\alpha(t)=C/\sqrt{t}$ for an appropriate C, the error scales as $L\sqrt{\frac{\min(d,n)\log(n^{1/2}dT)}{T(1-\max_k\sigma_2(P^k))}}$. 
The error scales as $T^{-1/2}$ which is a common factor we see in all results. 
The term $1/1-\sigma_2(P^k)$ determines how quickly nodes come to a consensus in coordinate $k$. When $P^k=P$, we do not obtain the factor $\min(d,n)$ retrieving the results of DDA \cite{duchi2012dual}.

\begin{lem}{\bf Round robin scheme:}
\label{lem:round robin scheme}
For the settings in Th. \ref{thm:basic_all}, the DCDA algorithm under the round robin $m-$coordinate sharing scheme is such that 
\ea{
& f(\hat{x}_i(T)) - f(x^*) \leq  \frac{\psi(x^*) }{T \alpha(T)} \\
& +\frac{L^2}{T}\sum_{t=1}^T  \alpha(t-1)\left(10 + \frac{12d\log 2\sqrt{n}T }{m(1- \sigma_2(P))}  \right).  \nonumber 
\label{eq:round robin scheme}
}
\end{lem}

With an appropriate choice of the step size, the error  in Lem. \ref{lem:round robin scheme} scales as $ L\sqrt{\frac{d\log(nT)}{mT(1-\sigma_2(P))}}$. Thus, we would need twice the amount of time to achieve a fixed error $\epsilon$ if we transmit half the number of coordinates $m$ at each time instant. 

\begin{lem}{\bf Randomized scheme:}
\label{lem: Randomized scheme}
%
For the settings in Th. \ref{thm:basic_all}, the DCDA algorithm under the randomized coordinate sharing scheme is such that, with probability greater than $1-\delta$
\begin{align}
& f(\hat{x}_i(T)) - f(x^*) \leq  \frac{\psi(x^*) }{T \alpha(T)} \\
& +\frac{L^2}{T}\sum_{t=1}^T  \alpha(t-1)\left(10 + 18\frac{\min(d,n) \log Tdn^{1/3}/\delta }{1- \max_k\sigma_2(\E[P^k(t)^2])}  \right). \nonumber 
\end{align}
\end{lem}

The result in Lem. \ref{lem: Randomized scheme} is similar to the static coordinate sharing case with the expected doubly stochastic sharing matrix used. Consider the specific case where the nodes collectively share coordinate $k$ with all other nodes with probability $\rho$. In this case, $P^k(t) = \frac{1}{n}\mb{1}\mb{1}^\intercal$ with probability $\rho$, else $P^{k}(t)=\mb{I}.$ In this case, the error scales as $L\sqrt{\frac{\log Tdn/\delta}{\rho T}}$ with high probability. Similar to the round robin case, the analysis shows an inverse dependence between the number of coordinates shared and the time needed for convergence.

\spaceup
\subsection{Variations of the DCDA algorithms}

In this section we study three variations of the DCDA scheme as introduced in Sec. \ref{sec:Problem formulation}. 
First, we consider the case in which 
each node does not have access to the exact gradient of its local function
but instead obtains a noisy estimate of this value. 
The DCDA algorithm for the stochastic gradient setting simply uses the stochastic gradient in place of the actual gradient. 
Convergence is studied under some mild assumptions on the noisy gradient value.

\begin{assumption}
\label{ass:stoch grad}
Assume $\mc{F}_t$ be the $\sigma$-field that contains all information known by all nodes till time $t$ and let  $g'(t)$ be the stochastic gradient at time $t$.
%
Further assume that:

\noindent
{\tiny $\bullet$} the stochastic gradient $g'_i(t)$ is an unbiased estimate of the actual gradient, i.e. $\E[g'_i(t)|\mc{F}_t]\in \partial f_i(x_i(t))$,

\noindent
{\tiny $\bullet$}
 the stochastic gradient is bounded. $\|g'_i(t)\|_*\leq  L$

\noindent
{\tiny $\bullet$}
The set $\mc{X}$ satisfies $\|x-x'\|\leq R ~\forall x,x'\in\mc{X}.$
\end{assumption}

\begin{lem}{\bf Stochastic gradient DCDA algorithm:}
\label{lem:Stochastic gradient DCDA algorithm}
For the settings in Th. \ref{thm:basic_all} and under the assumptions in Ass. \ref{ass:stoch grad}, the stochastic gradient DCDA algorithm is such that, with probability $1-\de$
\spaceup
\ea{
 f(\hat{x}_i(T)) - f(x^*)  \leq  \eqref{eq:basic_all} 
 +LR\sqrt{\frac{8\log\f 1 \delta}{T}}.
 \label{eq:Stochastic gradient DCDA algorithm}
 }
\end{lem}
From  Lem. \ref{lem:Stochastic gradient DCDA algorithm}, we conclude that the scaling of the error of the stochastic gradient DCDA algorithm is the same as the DCDA algorithm.

Let us next consider the noisy communication scenario. 
%
%
More precisely, message $z_j(t)$ transmitted at time $t$ from node $j$ to node $i$ suffers from additive noise $n_{ij}(t)$, i.e. $u_{ij}(t)=z_j(t) + n_{ij}(t)$ 
%

The DCDA algorithm for the noisy communication setting uses the noisy dual variable estimate $u_{ij}(t)$ instead of the actual value $z_j(t)$.
Convergence is shown under some assumptions on the noise sequence and for the static sharing scheme. 

\begin{lem}{\bf  Noisy communication static staring scheme DCDA algorithm:}
\label{lem:Noisy communication static coordinate scheme DCDA algorithm}
Consider the settings in Th. \ref{thm:basic_all} scheme where the function is $L$-Lipschitz with respect to the $\ell_2$-norm.
Further assume that there  exists $R$ such that $\sup_{x,x' \in \Xcal}||x-x'||_2\leq R$.
Under the assumptions that each $n_{ij}(t)$ has independent zero-mean sub-Gaussian components of power $\ga^2/d$, the noisy communication static sharing DCDA algorithm is such that, with probability $1-\de$
{\footnotesize \ea{
& f(\hat{x}_i(T))-f(x^*) \leq  \eqref{eq:static sharing}+ \ga (R+2L)\sqrt{\frac{2\log\f3 \delta}{nT}} + \sum_{t=1}^T \alpha(t-1)
\nonumber \\
&  \times\bigg(\frac{\ga^2(1+\sqrt{8} \log \f 3 {\delta} )}{ndT}  + \frac{3L}{T}\sqrt{\frac{2\ga^2\log \f {6Tnd} {\delta}}{(1-\max_k\sigma_2(P^{k})^2)}} \bigg). \label{eq:Noisy communication static coordinate scheme DCDA algorithm} 
}
}
\end{lem}

Finally, we consider the case in which the communication among nodes is quantized using infinite-level uniform quantization.
%
%
%
%
%
At each time step, a node broadcasts the quantized scaled dual variable update 
\ea{
[u_i(t)]_k= 
\bigg \lfloor  \f {[z_i(t)]_k-[z_i(t-1)]_k}{s(t)}  + u(t)\bigg \rfloor,
\label{eq:message quant}
}
 where $u(t)  \iid \Ucal([-1/2,+1/2])$  is dither used to guarantee that the quantization noise is uniformly distributed in the interval $[-1/2,+1/2]$, while 
 $s(t)>0$ is a zooming sequence that converges to zero and is known a priori to all nodes. 
%
The dual update operation is replaced by:
{\footnotesize\ea{
[z_i(t+1)]_k = \big[z_i(t)+g_i(t)-g_i(t-1) + \sum_j P^k_{ij}(t) s(t)u_j(t)\big]_k.  \label{eq:dual update quantized} 
}
}
\begin{lem}{\bf  Quantized communication static coordinate scheme DCDA algorithm:}
\label{lem:Quantized communication static coordinate scheme DCDA algorithm}
Consider the settings in Th. \ref{thm:basic_all} scheme where the function is $L$-Lipschitz with respect to the $\ell_2$-norm.
Furthermore define
\ea{
\nu(t)=\max_k\sum_{r=0}^{t}s^2(r)\sigma_2(P^k)^{2(t-r+1)}
}
Under these assumptions, the quantized communication static sharing DCDA is such that
{\footnotesize \ea{
& f(\hat{x}_i(T))-f(x^*) \leq  \eqref{eq:static sharing}+ 
R  \sqrt{\widehat{s^2}(T)\frac{\log1/\delta}{T}} +\sum_{t=1}^T \alpha(t-1)
\nonumber
\\
&  \times\bigg( \f {2s(t) L + s^2(t)}{nT}  +\frac{3L}{T}\sqrt{2 \nu(t)\log (2Tnd/\delta)} \bigg). \label{eq:Quantized communication static coordinate scheme DCDA algorithm}
%
}
}
%
\end{lem}
%
\section{Numerical Simulations}
\label{sec:Numerical Simulations}

For numerical simulations, we consider the scenario in which the function $f_i(x)$ arises from evaluating a common loss function $\ell$ over a set of $m$ local measurements $\{z_{ij}\}_{j=1}^m$: correspondingly we have
\begin{align}
F(x)=\sum_{i=1}^n f_i(x) = \sum_{i=1}^n \sum_{j=1}^m \ell(x,z_{ij}) \label{eq:optgoal local measurements}.
\end{align}
%

\begin{figure}
	\centering
	\includegraphics[width=0.9\linewidth]{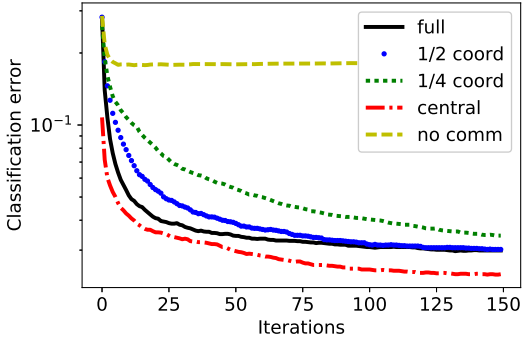}
    \vspace{-.25cm}
	\caption{Classification performance with distributed SVM \label{fig:svm}}	
    \vspace{-.5cm}
\end{figure}

%
%
\noindent
{\bf Support Vector Machine (SVM):}
In the first case, we look at using support vector machines for classification. Each data-point at a local node consists of a label $l_{ij}$ uniformly drawn from $\{-1,1\}$ and the data point $z_{ij}\stackrel{\mr{iid}}{\sim}\mathcal{N}(\mu_{l_{ij}},\Sigma)$.
The linear SVM algorithm finds a hyperplane that separates data drawn from the two distributions with the maximum margin
\ean{
\ell(x,(z_{ij}, l_{ij}))=\f 1 {2d} ||x||_2^2 
+C \sum_{j=1}^m \max \lb 1-l_{ij}x^T z_{ij}; 0 \rb.
}
In Fig. \ref{fig:svm} we plot the simulation results for $X\in \Rbb^{30}$, $n=10$ and $m=10$ and full network connectivity. 
At each instant in time, the nodes collectively sample a certain fraction $f$ of their coordinates to share.
For this scenario, we compare the performance for $f \in [0 \ 1/2 \ 1/4 \ 1]$.
The performance for the centralized SVM algorithm is also plotted for comparison.
%
As it can be observed, without any communication, nodes reach a suboptimal solution. To reach $90\%$ classification accuracy rate, nodes take twice as long if they share only half their coordinates. Finally, note the small loss in performance between the fully centralized scheme to the decentralized one. 

\begin{figure}
\includegraphics[width=0.9\linewidth]{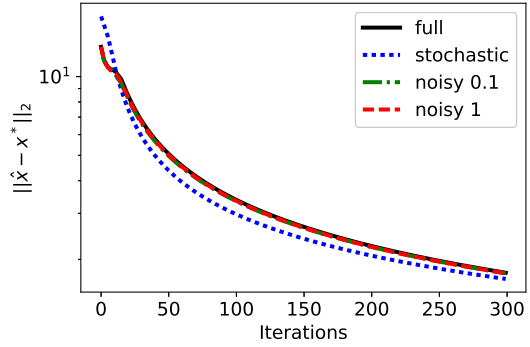}
\vspace{-.25cm}
\caption{Impact of stochastic gradients, noisy communications on performance of DCDA for linear regression \label{fig:lin_reg}}
\vspace{-.5cm}
\end{figure}

%
%
\noindent
{\bf Linear Regression:}
Next, we investigate the effect of noisy communication and stochastic gradient in the DCDA algorithm for the classic linear regression problem. 
We consider the case where $x\in\mathbb{R}^{30}$, $n=10$ and $m=20$ and a fully connected network. Note that local measurements consist of random normal measurement vectors $a_{ij}$ and the measurement $z_{ij}$.
%
For the noisy communication scenario, each node observes $z_{ij}=A_{ij} x +n_{ij}$,for $n_i \stackrel {\mr{iid}}{\sim}\mathcal{N}(0,\Sigma)$ and $f_i(x)=\ell(x,(A_{ij} ,z_{ij}))=\frac{1}{2}\|A_{i:} x-z_{i:}\|_2^2$ (with a slide abuse of notation). 
In the stochastic gradient case, nodes form mini-batches of size 4 as opposed to using all 20 data points for each iteration.
As can be seen in Fig.\ \ref{fig:lin_reg}, there is minimal loss in performance from using stochastic gradients or when additive noise is added to the nodes being shared. This suggests that cheaper computation using stochastic gradients, or quantization effects creating additive noise may not significantly alter performance. 

\begin{figure}
	\includegraphics[width=0.9\linewidth]{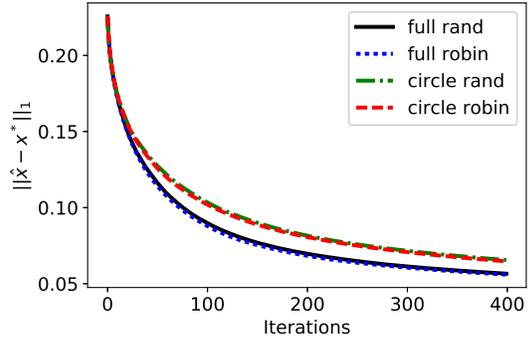}
\vspace{-.25cm}
\caption{Robust regression performance.}
\vspace{-.5cm}
	\label{fig:rob_reg}
\end{figure}

%
\noindent
{\bf Robust Regression:}
Finally, we consider the robust regression problem in which each node observes
$z_{ij}= A_{ij} x +(1-b_{ij}) o_{ij} + b_{ij}n_{ij}$ 
where $x$ is in the unit simplex ($[x]_i\geq 0,\|x\|_1=1$) , $b_{ij}$ is a binomial  noise that modulates between a large outlier Gaussian noise $o_{ij}$ or smaller additive Gaussian noise $n_{ij}$. The $\ell_1$ penalty is used as the loss function or $f_i(x)=\ell(x,(A_{ij} ,z_{ij}))=\|A_{i:} x-z_{i:}\|_1$.
%
%
For this problem, we consider an entropic proximal function that ensures that the estimate is in the probability simplex and  minimize the $\ell_1$-norm. 
In the simulation, we compare the round robin scheme and the randomized scheme where the amount of communication is kept equal. 
Both these schemes are compared for the fully connected network as well as a circle topology where nodes are connected to the closest neighbor on each side. 
Estimate $x \in \mathbb{R}^{20}$ and $m,n=10$.
%
The performance is presented in Fig. \ref{fig:rob_reg}. The performance of the fully connected layer is better than the circle topology because more communication is taking place, allowing estimates to quickly travel through the network. 
No significant  difference in performance between the randomized and round robin schemes is observed. 


\section{Conclusion}
\label{sec:Conclusion}
We considered the problem of distributed optimization with limited communication where nodes collectively solve a convex optimization problem  but have a limitation on the transmission among neighbors.
%
We proposed a distributed coordinate dual averaging algorithm for this problem and analyzed its performance. We showed that the time required to achieve a fixed accuracy would double if the rate limitation in messages between nodes were halved. 
We also show convergence in the scenario of stochastic gradients, noisy and quantized communication.
%

\newpage

\newpage

 \appendix
 

  

\newpage

\bibliographystyle{IEEEtran} 
\bibliography{references}

\newpage
\onecolumn
\begin{center}
    {\bf \Large APPENDIX}
\end{center}

\subsection*{Appendix A: Proof of Th. \ref{thm:basic_all}}

In this section prove the result in Th. \ref{thm:basic_all}  on the convergence of the DCDA algorithm.
In our  analysis of the DCDA algorithm, we employ techniques similar to those used in \cite{duchi2012dual} to study the convergence of the DDA algorithm. 

Before proceeding with the proof, let us introduce the auxiliary sequence $y(t)$ as the  projection  of $\zo(t)$ using the operator $\Pi_{\psi,\alpha(t)}$, that is
\eas{
	y(t) &= \Pi_{\psi,\alpha(t-1)}(\zo(t)).
	\label{eq: def yt}
}
An important consequence of  double symmetric stochastic nature of the matrices $P^k$ is a particularly simple evolution of the  variable $\zo(t)$ as
\ea{
	[\zo(t+1)]_k &= \frac{1}{n}\sum_{i=1}^n \sum_{j=1}^n P^k_{ij}[z_j(t)]_k + [g_i(t)]_k 
    \label{eq: dual evolution}\\
	&=   \frac{1}{n}\sum_{j=1}^n [z_j(t)]_k \sum_{i=1}^n  P^k_{ij}+\frac{1}{n}\sum_{i=1}^n  [g_i(t)]_k \nonumber  \\
	&= [\zo(t)]_k+ [\bar{g}(t)]_k. \nonumber 
}
As for the result in \cite[Th. 1]{duchi2012dual}, convergence of the local estimates to the global optimal is in terms of time averages $\hat{x}_i(t)$ and $\hat{y}(t)$
as Jensen's inequality can be used to show that
\ea{
f_i(\hat{x}_i(T)) &\leq \frac{1}{T}\sum_{t'=1}^T f_i(x_i(t)).
\label{eq:jensen}
}
Paralleling the analysis in \cite{duchi2012dual}, we bound the error estimate as
\eas{
	f(\hat{x}_i(T)) - f(x^*) &\leq f(\hat{y}(T))-f(x^*) + L\|\hat{x}_i(T) - \hat{y}(T) \| 
	\label{eq:i1 1}\\
	&\leq \frac{1}{Tn} \sum_{t=1}^T \sum_{j=1}^n f_j(x_j(t))-f_j(x^*)+ \frac{1}{Tn}\sum_{t=1}^T \sum_{j=1}^n  f(y(t)) - f_j(x_j(t)) + \frac{L}{T}\sum_{t=1}^T \|x_i(t) - y(t) \| 
 	\nonumber \\
 &\leq \frac{1}{Tn}\sum_{t=1}^T \sum_{j=1}^n  f_j(x_j(t))-f_j(x^*) + \frac{L}{Tn}\sum_{t=1}^T \sum_{j=1}^n \|y(t) - x_j(t)\| + \frac{L}{T}\sum_{t=1}^T\|x_i(t) - y(t) \| \label{eq:i1 11} \\
 	&\leq \frac{1}{Tn} \underbrace{\sum_{t=1}^T \sum_{j=1}^n  f_j(x_j(t))-f_j(x^*)}_{\Ga}  \label{eq:i1 2} \\
	& \quad \quad \quad \quad +  \frac{L}{Tn}\sum_{t=1}^T \sum_{j=1}^n \alpha(t-1)\|\zo(t) - z_j(t)\|_*  +\f L T\sum_{t=1}^T\alpha(t-1)\| \zo(t)-z_i(t)  \|_*, \nonumber
}{\label{eq:i1}}
where the inequality in \eqref{eq:i1 1}  follows  by the Lipschitz property of $f$.  
Equations \eqref{eq:i1 2}  and \eqref{eq:i1 11} follow  by the Lipschitz property of the projection operator as shown in the next lemma. 

\begin{lemma}{\bf \cite[Lem. 5]{duchi2012dual}}
	\label{lem:duallip}
	$\Pi_{\psi,\alpha}(z)$ is $\alpha$-Lipschitz with respect to the dual norm.
\end{lemma}

Next, we bound the term $\Ga$ in \eqref{eq:i1 2}:
%
\eas{
	\Ga
	&=\sum_{t=1}^T \sum_{j=1}^n f_j(x_j(t)) - f_j(x^*) \label{eq:Gamma1}\\
	&\leq \sum_{t=1}^T \sum_{i=1}^n \langle g_j(t), x_j(t)-x^*\rangle 
	\label{eq:Gamma2} \\
	&= n \underbrace{\sum_{t=1}^T  \langle \bar{g}(t),y(t)-x^*\rangle}_{\Psi_1} 
	\label{eq: y(t) comes in}\\
	& \quad \quad  \quad +\underbrace{\sum_{t=1}^T \sum_{i=1}^n \langle g_j(t), x_j(t)-y(t) \rangle}_{\Psi_2}, \label{eq: y(t) comes in_2}
}{\label{eq:bound gamma}}
where \eqref {eq:Gamma2} follows from the definition of subgradient. 
Let us first bound $\Psi_2$ in \eqref{eq: y(t) comes in_2}. As the function $f_j$ is L-Lipschitz, the gradient is bounded in the dual norm, i.e. $\|g_j\|_*\leq L$, the term $\langle g_j(t), x_j(t)-y(t) \rangle$ can be bounded using Cauchy–Schwarz inequality:
\eas{
\Psi_2 &\leq \sum_{t=1}^T \sum_{j=1}^n \|g_j(t)\|_* \|x_j(t)-y(t)\| \\
&\leq L \sum_{t=1}^T \sum_{j=1}^n \|\Pi_{\psi,\alpha(t-1)}(z_j(t))-\Pi_{\psi,\alpha(t-1)}(\zo(t))\| \nonumber\\
&\leq L \sum_{t=1}^T \sum_{j=1}^n  \alpha(t-1) \|\zo(t)-z_j(t)\|_*
\label{eq: last in},
}
where \eqref{eq: last in} follows from Lem. \ref{lem:duallip}.
%
%
%
%
%
Next, we bound the term  $\Psi_1$ in \eqref{eq: y(t) comes in}: 
%
this  portion  of the proof is similar to the bound in \cite[(20)-(21)]{duchi2012dual}. 
Starting from the  definition of $y(t)$, we write
\begin{align*}
	y(t) &= \Pi_{\psi,\alpha(t-1)}(\zo(t)) \\
	&= \argmin_{x\in\mathcal{X}} ~~ \langle \zo(t), x\rangle + \frac{1}{\alpha(t-1)}\psi(x) \\
	&= \argmax_{x\in\mathcal{X}}~~\langle- \zo(t), x\rangle - \frac{1}{\alpha(t-1)}\psi(x). 
\end{align*}
Let us next use the dual function $\psi^*_\alpha (z) = \sup_x \langle z,x \rangle - \frac{1}{\alpha}\psi(x)$ to find an upper bound $\Psi_1$.
Since the set $\mathcal{X}$ is closed and convex and $\psi$ is convex, we have that the supremum is attained. In other words, we have,
\begin{align}
	\psi^*_{\alpha(t-1)}(-\zo(t)) &= \langle -\zo(t), y(t) \rangle - \frac{1}{\alpha(t-1)}\psi(y(t)).
\end{align}
Accordingly,  for arbitrary $u\in\mathcal{X}$, we have
\begin{align*}
	\psi^*_{\alpha(t-1)}(u) &\geq \langle u, y(t) \rangle - \frac{1}{\alpha(t-1)}\psi(y(t)) \\
	&\geq \langle -\zo(t), y(t) \rangle - \frac{1}{\alpha(t-1)}\psi(y(t)) + \langle y(t), u+\zo(t)\rangle \\
	&= \psi^*_{\alpha(t-1)}(-\zo(t)) + \langle y(t), u+\zo(t)\rangle,
   \end{align*}
which shows that $\nabla\psi^*_{\alpha(t-1)}(-\zo(t))=y(t)$.
Next, from the strong convexity and positivity of $\psi$ we have that, for any $t\in[0,1]$, there exists $c\in[0,t]$ such that
\begin{align*}
\psi^*_{\alpha}(z-tg) &= \psi^*_{\alpha}(z) - t\langle \nabla\psi^*_{\alpha}(z-cg),g\rangle.
\end{align*}
Using $t=1$ and rearranging,
\ea{
	\psi^*_{\alpha}(z-g) &= \psi^*_{\alpha}(z) - \langle \nabla\psi^*_{\alpha}(z),g\rangle - \langle \nabla\psi^*_{\alpha}(z)-\nabla\psi^*_{\alpha}(z-cg),g\rangle 
	\nonumber \\
	&\leq \psi^*_{\alpha}(z) - \langle \nabla\psi^*_{\alpha}(z),g\rangle + \|g\|_*\|\nabla\psi^*_{\alpha}(z)-\nabla\psi^*_{\alpha}(z-cg)\| 
	\nonumber \\
	&\leq \psi^*_{\alpha}(z) - \langle \nabla\psi^*_{\alpha}(z),g\rangle + \alpha \|g\|_*^2.
	\label{eq:pp2}
}
Using \eqref{eq:pp2} for $z=\zo(t)$ and $g=\go(t)$ yields
%
%
\begin{align*}
	\psi^*_{\alpha(t)}(-\zo(t)-\bar{g}(t)) &\leq \psi^*_{\alpha(t-1)}(-\zo(t)-\bar{g}(t)) \\
	&\leq \psi^*_{\alpha(t-1)}(-\zo(t)) - \langle y(t), \bar{g}(t) \rangle + \alpha(t-1)\|\bar{g}(t)\|_*^2,
\end{align*}
and thus
\ea{
	\langle y(t), \bar{g}(t)\rangle &\leq \psi^*_{\alpha(t-1)}(-\zo(t)) - \psi^*_{\alpha_{t}}(-\zo(t+1)) + \alpha(t-1)\|\bar{g}(t)\|_*^2.
	\label{eq:11}
}
From the definition of the dual function, we have
\ea{
	\langle -\zo(T+1),x^*\rangle - \frac{1}{\alpha(t)}\psi(x^*) &\leq \psi^*_{\alpha(t)}(\zo(T+1)) \nonumber\\
	\Rightarrow \langle \zo(T+1), -x^* \rangle &\leq \frac{1}{\alpha(t)}\psi(x^*)  + \psi^*_{\alpha(t)}(\zo(T+1)).
	\label{eq:duality}
}
Combining \eqref{eq:duality} and \eqref{eq:11}, we obtain a bound on $\Psi$ in \eqref{eq: y(t) comes in} as
\ea{
	\Psi_1=\sum_{t=1}^T \langle \bar{g}(t),y(t)-x^*\rangle 
	& = \sum_{t=1}^T \lb \langle \bar{g}(t),y(t)\rangle+\langle \bar{g}(t),-x^*\rangle  \rb \nonumber \\
	&\leq 
	\sum_{t=1}^T  \alpha(t-1)\|\bar{g}(t)\|_*^2 + \frac{1}{\alpha(t)}\psi(x^*). 
	\label{eq:bound Psi}
}
Substituting the bound  in \eqref{eq:bound Psi} and \eqref{eq: last in}
in \eqref{eq:i1 2} yields the bound in \eqref{eq:basic_all}. 

\subsection*{Appendix B: Proof of Lem. \ref{lem:Static  sharing  scheme}}

In the static sharing scheme, each coordinate of the optimization variable follows a different but constant graph, $P^k(t) = P^k$,  for all $t \in \Nbb$.
In the analysis of this scheme, we rely on the result of Th. \ref{thm:basic_all} and bound the term $[\zo(t)-z_i(t)]_k$ using the properties of the communication protocol. 
To this end, we define
\ea{
\Phi^k(t,s)
& = P^k(t)P^k(t-1)\ldots P^k(s) \nonumber  \\
& =\prod_{j=t}^s P^k (j),
\label{eq:definition Phi k}
} 
so that the evolution of the dual variable can be expressed recursively as
\begin{align}
[z_i(t+1)]_k &= \sum_{j=1}^n [\Phi^k(t,s)]_{ij} [z_j(s)]_k +[g_i(t)]_k + \sum_{r=s+1}^{t} \sum_{j=1}^n  [\Phi^k(t,r)]_{ij}[g_j(r-1)]_k \label{eq:basicZupdate},
\end{align}
for any $0\leq s < t$.
We set the starting value $s=0,z_i(s)=0$. Now we can see
\ea{
[\zo(t)-z_i(t)]_k &= \sum_{r=1}^{t-1}\sum_{j=1}^n  \left( 1/n - [\Phi^k(t-1,r)]_{ij}\right)[g_j(r-1)]_k +\frac{1}{n}\sum_{j=1}^n  [g_j(t-1)-g_i(t-1)]_k.
}
In the following, we also resort to the following inequality from Perron-Frobenius theory for double-stochastic matrices: for any $x$ within the $n-$dimensional probability simplex:
\begin{align*}
\Big{\|} (P^k)^t x - \frac{\mb{1}}{n} \Big{\|}_1 &\leq  \sqrt{n} \Big{\|} (P^k)^t x - \frac{\mb{1}}{n} \Big{\|}_2 \leq \sqrt{n}\sigma_2(P^k)^t,
\end{align*}
where $\sigma_2( \cdot )$ is the second singular value of a matrix. 
We are now ready to bound the term $\|\zo(t)-z_i(t)\|_*$:
\eas{
\|\zo(t)-z_i(t)\|_* &\leq \sum_{r=1}^{t-1}\sum_{j=1}^n  \|g_j(r-1)\|_*\max_k \labs \f 1 n-[\Phi^k(t-1,r)]_{ij} \rabs+ \frac{1}{n}\sum_{j=1}^n  \|g_j(t-1)-g_i(t-1)\|_* \\
&\leq \sum_{r=1}^{t-1}L\sum_k \Big{\|}\frac{\mb{1}}{n}-\Phi^k(t-1,r)_i\Big{\|}_1 + 2L.
}{\label{eq:statice coordinate sharing abs(zo-z)}}
When $t-r\geq \Delta_t = \frac{\log \sqrt{n}dT/\epsilon}{-\log\min_k\sigma_2(P^k)}$, we have $\sum_k \|\frac{\mb{1}}{n}-\Phi^k(t-1,r)_i\|_1 \leq \frac{1}{T}$. Else, this sum would be less than $2\min(d,n)$. We see this as,
\ean{
\sum_{j=1}^n  \max_k \labs \frac{1}{n}-[\Phi^k(t-1,r)]_{ij} \rabs &\leq \sum_k \Big{\|}\frac{\mb{1}}{n}-\Phi^k(t-1,r)_i\Big{\|}_1 \\
&= \sum_k \left \|\frac{\mb{1}}{n} \right\|_1 + \left\|\Phi^k(t-1,r)_i \right\|_1 = 2d,
}
and
\ean{
\sum_{j=1}^n  \max_k \labs \frac{1}{n}-[\Phi^k(t-1,r)]_{ij} \rabs &\leq \sum_{j=1}^n  \frac{1}{n}+\max_k [\Phi^k(t-1,r)]_{ij} = 2n.
}
Thus we can bound the sum 

\eas{
\|\zo(t)-z_i(t)\|_* &\leq \sum_{r=t-\Delta_t}^{t-1} L\sum_k \Big{\|}\frac{\mb{1}}{n}-\Phi^k(t-1,r)_i\Big{\|}_1 + \sum_{r=1}^{t-\Delta_t}\sum_k \Big{\|}\frac{\mb{1}}{n}-\Phi^k(t-1,r)_i\Big{\|}_1 + 2L \\
&\leq 2\min(d,n)L\frac{\log \sqrt{n}dT/\epsilon}{-\log\min_k\sigma_2(P^k)} + L + 2L \\
& = L \lb 2\min(d,n) \frac{\log \sqrt{n}dT/\epsilon}{-\log\min_k\sigma_2(P^k)} + 3  \rb
\label{eq:static}.
}

\begin{remark}
Note the factor of $\min(d,n)$ in \eqref{eq:static}: this can be reduced if multiple coordinates follow the same weight function. In an extreme case, this factor of $d$ would be removed if all coordinates are transmitted at each point in time. 
\end{remark}

\subsection*{Appendix C: Proof of Lem. \ref{lem:round robin scheme}}

In the analysis round robin scheme, $m$ of of the $d$ components of the dual vector are shared at each point in time. 
%
Accordingly, any given coordinate is shared among a set of nodes every $\kappa=d/m$ time instances. 
For simplicity, assume that $m$ divides $d$: if this is not the case, one can upper bound performance considering $m'=d \lfloor m \ d\rfloor$.
%
%

At the time instant $T=l\ka+\tau $, the coordinates $[k_{ m+1}\ldots k_{\tau(m+1)}]$ are transmitted by each node over 
the graph $P$.
Using the notation in \eqref{eq:basicZupdate} and given that $m$ divides $d$, we write
\begin{align*}
\Phi^{k_h}(T,r) &= P^{k_h}(T)\cdots P^{k_h}(r) \\
&= \begin{cases}
P^{\lfloor \frac{T-r}{\kappa}\rfloor} & r< T - \left( \lfloor \frac{T-r}{\kappa}\rfloor +1 \right)\kappa + \tau \\
P^{\lfloor \frac{T-r}{\kappa} \rfloor+1} & \textrm{otherwise}.
\end{cases}
\end{align*}
We can now see that
\eas{
\|\zo(T+1)-z_i(T+1)\|_* &\leq \sum_{r=1}^{T}\sum_{j=1}^n  \|g_j(r-1)\|_*\max_{h} |1/n-[\Phi^{k_h}(T,r)]_{ij}| + \frac{1}{n}\sum_{j=1}^n  \|g_j(T)-g_i(T)\|_* \\
&\leq \sum_{r=1}^{T}\sum_{j=1}^n  \|g_j(r-1)\|_* \left( |1/n-[P^{\lfloor \frac{T-r}{\kappa}\rfloor}]_{ij}| + |1/n-[P^{\lfloor \frac{T-r}{\kappa}\rfloor+1}]_{ij}|\right)+ \frac{1}{n}\sum_{j=1}^n  \|g_j(T)-g_i(T)\|_* \nonumber \\
&\leq L\sum_{r=1}^{T} \|1/n-[P^{\lfloor \frac{T-r}{\kappa}\rfloor}]_{i} \|_1 + \|1/n-[P^{\lfloor \frac{T-r}{\kappa}\rfloor+1}]_{i}\|_1 + 2L.
\label{eq:sum rr 1}
}{\label{eq:sum rr}}
When $r<T-\Delta_t\kappa$ where $\Delta_t = \frac{\log 2\sqrt{n}T}{-\log \sigma_2(P)}$, we have
\begin{align*}
\|1/n-[P^{\lfloor \frac{T-r}{\kappa}\rfloor}]_{i} \|_1 + \|1/n-[P^{\lfloor \frac{T-r}{\kappa}\rfloor+1}]_{i}\|_1 &\leq \frac{1}{T},
\end{align*}
else, 
\ea{
\|1/n-[P^{\lfloor \frac{T-r}{\kappa}\rfloor}]_{i} \|_1 + \|1/n-[P^{\lfloor \frac{T-r}{\kappa}\rfloor+1}]_{i}\|_1 &\leq 4.
\label{eq:rr last}
}
%
%
Thus the summation in \eqref{eq:sum rr 1} is further loosened as 
\ea{
\|\zo(T+1)-z_i(T+1)\|_* &\leq 4L \frac{d\log 2\sqrt{n}T }{-m\log \sigma_2(P)} + 3L.
}
Recall also that $\log x \leq x-1$ for $x\in(0,1)$. Hence $1/-\log\sigma_2(P) \leq 1/1-\sigma_2(P)$. Combining this observation with  \eqref{eq:rr last} yields the bound in \eqref{eq:round robin scheme}.





\subsection*{Appendix D: Proof of Lem. \ref{lem: Randomized scheme}}
%
%
In the randomized communication scheme, nodes select a different subset of coordinates to transmit at each time step.
This selection is assumed to be  random and independent. 
%
%
This communication strategy can be implemented, for instance, by having nodes share a random seed and hence share the same random subset of coordinates. 
Another alternative is the one in which  nodes do not share a random seed and decide at each time step independent of each other what coordinates to transmit. 

We first prove the convergence of the random time-varying $\Phi(t,s)$ to $\mb{1}\mb{1}^{\intercal}/n$: consider a $u(t)$ in the probability simplex evolving as $u(t+1)=P^k(t)u(t)$, then 
\ea{
\E [\langle u(t+1)-\mb{1}/n,u(t+1)-\mb{1}/n \rangle | u(t)] &= (u(t)-\mb{1}/n)^\intercal \E[P^k(t)^2](u(t)-\mb{1}/n) \nonumber \\
&\leq \| u(t) - \mb{1}/n \|_2^2 \lambda_2(\E[P^k(t)^2]),
\label{eq:random 1}
}
where,  \eqref{eq:random 1} follows  because the leading eigenvalue and eigenvector of $\E[P^k(t)^2]$ are  $1$ and $\mb{1}/n$ respectively. 
Next, Chebyshev inequality can be used to derive the bound
\begin{align*}
\Pr \left( \|u(t)-\mb{1}/n\|_2 \geq \epsilon | u(0)\right) &\leq \frac{\E[|u(t)-\mb{1}/n\|_2 | u(0)]}{\epsilon^2} \\
&\leq \epsilon^{-2}\|u(0)-\mb{1}/n\|_2^2 \E[P^k(t)^2]^t,
\end{align*}
and thus we have
\begin{align*}
\Pr \left( \|\Phi^k(t,s)e_i - \mb{1}/n\|_2 \geq \epsilon \right) &\leq \epsilon^{-2}\lambda_2(\E[P^k(t)^2])^{t-s+1}.
\end{align*}
Now, with $t-s>\Delta_k=\frac{\log T^3d^3n/\delta}{1-\lambda_2(\E[P^k(t)^2])}$, we have that,with probability $1-\frac{\delta}{Td}$
\eas{
\left \| \frac{\mb{1}}{n}-\Phi^k(t-1,s)e_i \right\|_1 &\leq \sqrt{n}\left\| \frac{\mb{1}}{n}-\Phi^k(t-1,s)e_i\right\|_2 
\\
&\leq \sqrt{n} \left\|\frac{\mb{1}}{n}-\Phi^k(t-1,t-\Delta_k)e_i\right\|_2 
\label{eq: randomized 2}\\
&\leq \frac{1}{Td}, \nonumber
}{\label{eq: randomized 2 1/n}}
where the inequality in \eqref{eq: randomized 2} follow  because, for any $s'<s$, $\Phi(t-1,s')=\Phi(s,s')\Phi(t-1,s)$ , $\|\Phi(s,s')\|_2\leq 1$, and  $\|\Phi(t-1,s')e_i-\mb{1}/n\|_2\leq \|\Phi(s,s')\|_2\|\Phi(t-1,s)e_i-\mb{1}/n\|_2$. 
Note that for $k\neq k'$, $\E[P^k(t)^2]=\E[P^{k'}(t)^2]$, so that the bound in \eqref{eq: randomized 2 1/n} holds for any $k$ and and the same $\Delta_k=\Delta_{k'}=\Delta$.

\subsection*{Appendix E: Proof of Lem. \ref{lem:Stochastic gradient DCDA algorithm}}
%

In a stochastic gradient DCDA algorithm, each node does not have access to the gradient of its local function. For instance, if a node may use a fraction of its total data to compute a gradient. This provides a cheap but noisy estimate of the gradient.
%
%
For this scenarios, we are again concerned with the convergence analysis as in Th. \ref{thm:basic_all}.
For this reason, we consider the same bounding as in  \eqref{eq:i1} and consider the additional terms arising from the presence of a stochastic gradient.

The difference in analysis with respect to the derivation in App. A starts from $\Gamma$ in Eq.\ \eqref{eq:Gamma1}:
\ea{
\Gamma &= \sum_{t=1}^T\sum_{i=1}^n  f_i(x_i(t)) - f_i(x^*) 
\label{eq:bound gamma stoch}\\
&\leq \sum_{t=1}^T\sum_{i=1}^n  \langle g_i'(t), x_i(t)-x^* \rangle \nonumber  \\
&\leq \sum_{t=1}^T\sum_{i=1}^n  \langle g_i(t), x_i(t)-x^* \rangle + \sum_{t=1}^T\sum_{i=1}^n  \langle g_i(t)-g'_i(t), x_i(t)-x^* \rangle. \nonumber 
}
The analysis of the first term proceed as in the  analysis of of \eqref{eq:Gamma2}  in App. A. The second term forms a martingale sequence as 
\begin{align*}
\E[\langle g_i(t)-g'_i(t), x_i(t)-x^* \rangle|\mc{F}_{t-1}] &= \langle g_i(t)-\E[g'_i(t)|\mc{F}_{t-1}], x_i(t)-x^* \rangle =0,
\end{align*}
and also bounded as
\ea{
\langle g_i(t)-g'_i(t), x_i(t)-x^* \rangle &\leq 2LR
}
give the assumption on the boundedness of $\|x-x'\|$ in Ass. \ref{ass:stoch grad}.
As $\langle g_i(t)-g'_i(t), x_i(t)-x^* \rangle$ is a bounded martingale sequence, we can apply the Azuma-Hoeffding inequality to obtain
\begin{align*}
\Pr \lsb \sum_{t=1}^T\sum_{i=1}^n  \langle g_i(t)-g'_i(t), x_i(t)-x^* \rangle \geq Tn\epsilon \rsb &\leq \exp \left(-\frac{T\epsilon^2}{8L^2R^2} \right).
\end{align*}
Thus, with probability $1-\delta$, we have
\ea{
\frac{1}{nT} \sum_{t=1}^T\sum_{i=1}^n  \langle g_i(t)-g'_i(t), x_i(t)-x^* \rangle &\leq LR\sqrt{\frac{-8\log\delta}{T}}.
\label{eq: azuma}
}
The remaining terms in \eqref{eq:i1} are bounded as in App. A, so that the 
inequality in \eqref{eq:Stochastic gradient DCDA algorithm} is obtained as \eqref{eq:basic_all} plus the additional term in  \eqref{eq: azuma}.

\subsection*{Appendix F: Proof of Lem. \ref{lem:Noisy communication static coordinate scheme DCDA algorithm}}

In this appendix, we prove convergence of the  noisy communication  DCDA algorithm for the static staring communication strategy.
The proof is an adaptation of the proof of Lem. \ref{lem:Static  sharing  scheme} that accounts for the additional additive noise term. 
Similarly to the derivation in App. E, the result in Th. \ref{thm:basic_all} can be leveraged to study the scenario of noisy communications in two steps: (i) bound  the additional terms arising in $\Gamma$ because of the presence of the additive noise as in \eqref{eq:bound gamma stoch}, and (ii) bound the convergence of the term $\|\zo(t)-z_i(t) \|_*$ under the assumption it the statement of Lem. \ref{lem:Noisy communication static coordinate scheme DCDA algorithm} on the choice of norm and communication startegy. 

Let us begin from the bounding of $\|\zo(t)-z_i(t) \|_*$: note that the update of the dual variable $z_i$ is now obtained as
%
%
\eas{
[z_i(t+1)]_k &= \sum_{j=1}^n  P^k_{ij}(t)\left[z_j(t) + n_{ij}(t)\right]_k + [g_i(t)]_k \\
&= \sum_{j=1}^n  P^k_{ij}(t)\left[z_j^p(t)+n_j(t)+n_{ij}(t)\right]_k +  [g_i(t)]_k    \label{eq: noisy 1} \\
&= \sum_{j=1}^n  P^k_{ij}(t)[z_j(t)^p]_k + [g_i(t)]_k + \sum_{j=1}^n   P_{ij}^k \lsb  n_j(t)+n_{ij}(t)]_k \rsb \nonumber \\
& = [z^p_i(t+1)]_k+[n_i(t+1)]_k,
\label{eq: noisy 2}
}
where, in \eqref{eq: noisy 1} we define $z_j^p(t)$ as the dual variable evolution in the noiseless case as in  \eqref{eq: dual evolution} (save for the spacial average), while 
$n_i(t)$ is defined the total accumulated noise on the dual variable $z_i(t)$, through the recursion 
\ea{
[n_i(t+1)]_k=\sum_{j=1}^n  P_{ij}^k \lsb n_j(t)_k+ n_{ij}(t) \rsb.
\label{eq:recursion 1}
}
Note that the recursion in \eqref{eq:recursion 1} can be reformulated using the notation in 
\eqref{eq:definition Phi k} as
\ea{
[n_i(t+1)]_k=\sum_{r=1}^{t+1}\sum_{j=1}^n  \Phi^k_{ij}(t+1,r)[n_{ij}(r-1)]_k.
\label{eq:recursion 2}
}
Under the assumption that the noise $n_{ij}(t)$ has independent zero-mean sub-Gaussian components (can be weakened to rotationally symmetric vector with sub-Gaussian tails) of power $\ga^2/d$. 
Additionally, 
\eas{
\E[z_i(t)-z^p_i(t)] & =0 \\
\E[\zo(t) - z_i(t) - \zo^p(t)+z^p_i(t)] &=0.
}
%
%
Next, let us define $\Delta z_i(t) = \zo(t)-z_i(t)$ and  $\Delta z_i^p(t) = \zo^p(t)-z_i^p(t)$  so that their difference,  $\Delta z_i(t+1)-\Delta z^p_i(t+1)$ is also sub-Gaussian and can be expressed as 
\ea{
[\Delta z_i(t+1)-\Delta z^p_i(t+1)]_k &= \sum_{r=1}^{t+1}\sum_{j=1}^n  \left( \frac{1}{n}-\Phi^k_{ij}(t+1,r) \right)[n_{ij}(r-1)]_k,
}
so that the variance proxy  $\sgs( \cdot)$ is obtained as 
\ea{
\sgs\left([\Delta z_i(t+1)-\Delta z^p_i(t+1)]_k\right) &\leq  \frac{\ga^2}{d}\sum_{r=1}^{t+1}  \left\| \Phi^k_{i}(t+1,r) -\frac{\mb{1}}{n}\right\|_2^2 \nonumber \\
&\leq \frac{\ga^2}{d(1-\sigma_2(P^k)^2)}, 
\label{eq:delta z-zp 2}
}
where, in \eqref{eq:delta z-zp 2},  we have used the fact that $\|\Phi^k_{ij}(t,r)-\mb{1}/n\|_2 = \|(P^k)^{t-r+1}e_i - \mb{1}/n\|_2 \leq \sigma_2(P^k)^{t-r+1}$. 
Using the deviation bound for sub-Gaussian variables
\ea{
\Pr(|X-\E[X]|\geq \epsilon)\leq 2\exp(-\epsilon^2/2 \sgs),
}for $X$ sub-Gaussian with  with variance proxy  parameter $\sgs$, we conclude that,  with probability $1-\delta$
\ea{
|[\Delta z_i(t)-\Delta z^p_i(t)]_k| &\leq \ga \sqrt{\frac{2\log 2Tdn/\delta}{d(1-\sigma_2(P^k)^2)}} \quad \forall \ t\leq T, i\in [n], k \in [d] 
\label{eq:delta z-zp 2}.
}
Thus with probability $1-\delta$, we have
\begin{align}
\frac{1}{T} \sum_{t=1}^T \|\zo(t)-z_i(t) \|_* &\leq \frac{1}{T} \sum_{t=1}^T \|\zo^p(t)-z^p_i(t) \|_* +  \frac{1}{T}\sqrt{\frac{2\ga^2\log 2Tnd/\delta}{d(1-\max_k\sigma_2(P^{k})^2)}}\|\mb{1}\|_*,
\label{eq:penalty 1}
\end{align}

\medskip

Let us next return to the bounding of $\Ga$ as in \eqref{eq: y(t) comes in}: in the 
  scenario of noisy communication this term can be bounded similarly to the stochastic gradient scenario by letting $g_i'(t)=g_i(t)+\sum_j P^k_{ij}(t)n_{ij}(t)$.
With this definition, the term $\Psi_1$ in  \eqref{eq: y(t) comes in} can be bounded as
\eas{
& \sum_{t=1}^T  \langle \overline{g'}(t),y(t)-x^*\rangle \\
&= \sum_{t=1}^T  \langle \bar{g}(t) + \bar{n}(t),y(t)-x^*\rangle - \sum_{t=1}^T  \langle \bar{n}(t),y(t)-x^*\rangle \\
& \leq \sum_{t=1}^T  \alpha(t-1)\left(\|\bar{g}(t) + \bar{n}(t)\|_*^2\right)+ \frac{1}{\alpha(T)}\psi(x^*) - \sum_{t=1}^T  \langle \bar{n}(t),y(t)-x^*\rangle. \\
& \leq \sum_{t=1}^T  \alpha(t-1)\left(\|\bar{g}(t)\|_*^2 + \|\bar{n}(t)\|_*^2\right)+ \frac{1}{\alpha(T)}\psi(x^*) + \sum_{t=1}^T  \langle \bar{n}(t),x^*-y(t) + 2\alpha(t-1)\bar{g}(t)\rangle. 
}{\label{eq:Ga bound for noisy}}
Here, $\bar{n}(t) = \frac{1}{n}\sum_{i}\sum_j P^k_{ij}(t)n_{ij}(t)$. To proceed in the bounding of the term $\sum_{t=1}^T \alpha(t-1)\|\bar{n}(t)\|_*^2$, we utilize the assumption that $\|\cdot\|$ is the $\ell_2$-norm and that $\sup_{x,x'}\|x-x'\|\leq R$. 
Since  $[\bar{n}(t)]_k\sim \mc{N}(0,\frac{\ga^2}{dn})$, we have that  $\|\bar{n}(t)\|_2^2$ is a sub-exponential variable with parameters $(\frac{2\ga^2}{d\sqrt{n}},\frac{4\ga^2}{nd})$. 
Accordingly,  $\sum_{t=1}^T \alpha(t-1)\|\bar{n}(t)\|_2^2$ is sub-exponential with parameters \ean{
\lb \frac{2\ga^2}{d\sqrt{}n}\sqrt{ \sum_{t=1}^T \alpha^2(t-1)},\max_{t>1} \ \ \frac{4\ga^2\alpha(t)}{nd} \rb.
}

%
%
We now employ the concentration inequality for sub-exponential random variable $Z$ with parameters $(\nu,b)$:
\begin{align*}
\Pr(Z\geq \E[Z]+\epsilon) &\leq \exp \left( -\frac{\epsilon^2}{2\nu^2} \right) ~~\forall \epsilon\leq \frac{\nu^2}{b}.
\end{align*} 

We get that with probability greater than $1-\delta$ and large enough $T$ is 
\begin{align}
\frac{1}{T}\sum_{t=1}^{T}\alpha(t-1)\|\bar{n}(t)\|_2^2 &\leq \frac{\ga^2}{ndT}\left\{\sqrt{8 \sum_{t=1}^T\alpha^2(t-1)}\log1/\delta + \sum_{t=1}^T \alpha(t-1) \right\}  \nonumber \\
&\leq \frac{\ga^2}{ndT}\lb \sqrt{8}\log\f 1 \delta +  1\rb \sum_{t=1}^T \alpha(t-1).
\label{eq:penalty 2}
\end{align}
We can see that $\langle \bar{n}(t),x^*-y(t)+2\alpha(t-1)\bar{g}(t)\rangle$ is sub-Gaussian with zero mean and n 
variance proxy parameter
\ean{
\sgs(\langle \bar{n}(t),x^*-y(t)+2\alpha(t-1)\bar{g}(t)\rangle) &\leq \frac{\ga^2(R+2L)^2}{n},
}
so that, with probability $1-\de$ 
\ea{
\labs \frac{1}{T} \sum_{t=1}^T \langle \bar{n}(t),x^*-y(t)+\alpha(t-1)\bar{g}(t)\rangle \rabs &\leq \ga (R+2L)\sqrt{\frac{2\log \f 1 \delta}{nT}}.
\label{eq:penalty 3}
}
Combining the penalties in \eqref{eq:penalty 1}, \eqref{eq:penalty 2} and \eqref{eq:penalty 3} we obtain the bound  in \eqref{eq:Noisy communication static coordinate scheme DCDA algorithm}.
%
%
%

\subsection*{Appendix G: Proof of Lem. \ref{lem:Quantized communication static coordinate scheme DCDA algorithm}}

We finally come to the analysis of the performance of the DCDA algorithm with quantized communication in the static sharing scheme.  
Similarly to the derivation in \eqref{eq: noisy 2} in App. F for the noisy communication  scenario, the dual variable update can be rewritten as
\eas{
[z_i(t+1)]_k &= [z_i(t)]_k+\sum_{j \in N^k(i)} P_{ij}^k [(z_j(t)-z_j(t-1))+s(t)\De_j(k)]_k + [g_i(t)]_k-[g_i(t-1)]_k
\label{eq:quant dual 1}\\
& = \sum_{j \in N^k(i)} P_{ij}^k [z_j(t-1)]_k+\sum_{j \in N^k(i)} P_{ij}^k [(z_j(t)-z_j(t-1))+s(t)\De_j(k)]_k + [g_i(t)]_k \nonumber \\
& = \sum_{j \in N^k(i)} P_{ij}^k [z_j(t)]_k + [g_i(t)]_k + s(t) \sum_{j \in N^k(i)} P_{ij}^k\De_j(k) \nonumber \\
& =  z^p_i(t+1)]_k+ n_i(t+1),
\label{eq:quant dual 2}
}
where, in \eqref{eq:quant dual 1} we have used the definition of the transmitted message and the dual variable update in \eqref{eq:message quant} and \eqref{eq:dual update quantized} respective, and where we have defined $\De_j(t)$ as the quantization error, i.e.
\ea{
[\De_j(t)]_k=u_i(t)]_k- \bigg \lfloor  \f {[z_j(t)]_k-[z_j(t-1)]_k}{s(t)}   
\bigg \rfloor,
}
while, in \eqref{eq:quant dual 2}, we have defined $z_i^p(t)$ as in App. F  as the value of the dual variable in the noiseless case. Also, similarly to  App. F, $n_i(t)$ as the total accumulated error between $z_i^p(t)$ and $z_i(t)$, obtained as
\ea{
[n_i(t+1)]_k=\sum_{j=1}^n  P_{ij}^k \lsb n_i(t)_k+ \De_j(t) \rsb.
\label{eq:recursion 1 quant}
}
Similarly to \eqref{eq:recursion 2}, $n_i(t+1)$ can be recursively expressed as 
\ea{
[n_i(t+1)]_k=\sum_{r=1}^{t+1}\sum_{j=1}^n  \Phi^k_{ij}(t+1,r)[\De_j(r-1)]_k.
\label{eq:recursion 2 quant}
}
The expression in  \eqref{eq:recursion 2 quant} is  similar to the noisy communications  scenario in App. F, except for the fact that the noise has reducing variance (due to the zoom in sequence). 

Following the prior analysis, 
\begin{align*}
[\Delta z_i(t+1)-\Delta z^p_i(t+1)]_k &= \sum_{r=1}^{t+1}\sum_{j=1}^n  \left( \frac{1}{n} - \Phi^k_{ij}(t+1,r) \right)s(r-1)[\Delta_j(r-1)]_k 
\end{align*}
Each $s(t)\Delta_i(t)$ is bounded and hence sub-Gaussian with variance proxy parameter $s^2(t)$. 
Thus the above difference is sub-Gaussian with parameter
\begin{align*}
\sgs([\Delta z_i(t+1)-\Delta z^p_i(t+1)]_k) &\leq \sum_{r=1}^{t+1} s^2(r-1)\left\|\frac{1}{n}-\Phi^k_i(t+1,r) \right\|_2^2 \\
&\leq \sum_{r=1}^{t+1} s^2(r-1)\sigma_2(P^k)^{2(t-r+1)}
\end{align*}

Thus with probability greater than $1-\delta$, we have
\begin{align}
\frac{1}{T} \sum_{t=1}^T \|\zo(t)-z_i(t) \|_* &\leq \frac{1}{T} \sum_{t=1}^T \|\zo^p(t)-z^p_i(t) \|_* +  \frac{1}{T}\sqrt{2\log (2Tnd/\delta)\max_k\sum_{r=0}^{t}s^2(r)\sigma_2(P^k)^{2(t-r+1)}}\|\mb{1}\|_*%
\label{eq:bound zo-z quant}
\end{align}
As in App. F, we next note that the bounding of $\Ga$ as in \eqref{eq:bound gamma stoch} can be adapted to the case of quantized communication. By again letting $g_i'(t)=g_i(t)+\sum_j \Delta_i(t)$, we write
\eas{
	\Ga & \leq \sum_{t=1}^T\sum_{i=1}^n  \langle g'_i(t), x_i(t)-x^* \rangle + \sum_{t=1}^T\sum_{i=1}^n  \langle g_i(t)-g'_i(t), x_i(t)-x^* \rangle
	\label{eq:gamma quant} \\
	& \leq n \lb 	\sum_{t=1}^T  \langle \bar{g}'(t),y(t)-x^*\rangle  - \langle \bar{\Delta}(t),y(t)-x^*\rangle \rb.
	}
For the first term in \eqref{eq:gamma quant}, we have
 \ea{
    \sum_{t=1}^T  \langle \bar{g}'(t),y(t)-x^*\rangle &\leq \sum_{t=1}^T  \alpha(t-1)\left(\|\bar{g}(t)+\bar{\Delta}(t)\|_*^2\right)+ \frac{1}{\alpha(T)}\psi(x^*)
    \label{eq: go quant}.
  }
Considering the case of the $\ell_2$-norm, we get that the first term in \eqref{eq: go quant} is bounded by 
\begin{align*}
\sum_{t=1}^T  \alpha(t-1)\left(\|\bar{g}(t)+\bar{\Delta}(t)\|_*^2\right) &\leq \sum_{t=1}^T \alpha(t-1)(L^2+2s(t)\|1\|_2L + s^2(t)) 
\end{align*}

Finally, $\langle\bar{\Delta}(t), y(t)-x^* \rangle$ is zero-mean and bounded as $s(t)\|1\|_2R.$ Applying the Azuma-Hoeffding inequality, we get with probability greater than $1-\delta$
\ea{
\frac{1}{T} \sum_{t=1}^T \langle \bar{\Delta}(t), y(t)-x^*\rangle &\leq R\|1\|_2\sqrt{\widehat{s^2}(T)\frac{\log1/\delta}{T}}
\label{eq:bound 3}
}
Combining the bounds in \eqref{eq:bound zo-z quant}, \eqref{eq: go quant} and  \eqref{eq:bound 3} yields \eqref{eq:Quantized communication static coordinate scheme DCDA algorithm}.

\end{document}

%% file: header.tex
\newcommand{\argmax}{\operatorname{argmax}}
\newcommand{\argmin}{\operatorname{argmin}}

\newcommand{\lb}{\left(}
\newcommand{\rb}{\right)}
\newcommand{\lsb}{\left[}
\newcommand{\rsb}{\right]}

\usepackage[utf8]{inputenc}
\usepackage[tight,footnotesize]{subfigure}
\usepackage[cmex10]{amsmath}
\usepackage{amssymb}
\usepackage{amsthm}
\usepackage{wasysym}
\usepackage{balance}
\usepackage{url}

\usepackage{graphicx}
\usepackage{caption}
\usepackage{epstopdf}
\usepackage{tikz}
\usepackage{tikz-3dplot}
\usetikzlibrary{fit}
\usepackage{pgfplots}
 \usepgfplotslibrary{groupplots}
 \usepgfplotslibrary{external}
\usepackage{textcomp}

\usepackage{color}
\usepackage{bm}

\newtheorem{lemma}{Lemma}
\newtheorem{thm}{Theorem}

\newcommand{\be}{\begin{equation}}

\newcommand{\bc}{\begin{center}}
\newcommand{\bfl}{\begin{flushleft}}

\newcommand{\beqa}{\begin{eqnarray}}
\newcommand{\beqan}{\begin{eqnarray*}}
\newcommand{\beq}{\begin{equation}}

\newcommand{\E}{{\mathbb {E}}}

\newcommand{\mc}{\mathcal}
\newcommand{\mb}{\mathbf}

\newcommand{\mr}{\mathrm}

\newcommand{\iid}{\stackrel{\mathrm{iid}}{\sim}}

\renewcommand{\Pr}{\mathrm{Pr}}

\renewcommand*\env@matrix[1][*\c@MaxMatrixCols c]{%
  \hskip -\arraycolsep
  \let\@ifnextchar\new@ifnextchar
  \array{#1}}
  
\usepackage{etoolbox}
\preto{\section}{\setcounter{subsubsection}{0}}